\documentclass{aa}
\usepackage{psfig}

\def\sq{\hbox{\rlap{$\sqcap$}$\sqcup$}}
\newcommand{\be}{\begin{eqnarray}}
\newcommand{\ee}{\end{eqnarray}}

\begin{document}

\title{Gauss--Legendre Sky Pixelization (GLESP) for CMB maps}

\author{
A.G.\,Doroshkevich$^{1,2}$,
P.D.\,Naselsky$^{1,3,4}$,
O.V.\,Verkhodanov$^{1,5}$,
D.I.\,Novikov$^{2,6}$,
V.I.\,Turchaninov$^{7}$,
I.D.\,Novikov$^{1,2,3,8}$,
P.R.\,Christensen$^{1,3}$,
L.-Y.\,Chiang$^{1,3}$
}

\institute{
      Theoretical Astrophysics Center, Juliane Maries Vej 30, DK-2100,
      Copenhagen, Denmark  
\and  Astro Space Center of Lebedev Physical Institute, Profsoyuznaya 84/32,
      Moscow, Russia   
\and  Niels Bohr Institute, Blegdamsvej 17, DK-2100 Copenhagen,
      Denmark  
\and  Rostov State University, Zorge 5, Rostov-Don, 344090, Russia
\and  Special Astrophysical Observatory, Nizhnij Arkhyz, Karachaj-Cherkesia,
      369167, Russia
\and  Institute for Teoretisk Astrofysikk, Universitetet i Oslo, Postboks
      1029 Blindern 0315, Oslo, Norway
\and  Keldysh Institute of Applied Math, Russian Academy of Science,
      125047 Moscow, Russia
\and  University Observatory, Juliane Maries Vej 30, DK-2100, Copenhagen,
      Denmark
 }

\date{Received / Accepted}

\abstract{
A new scheme of sky pixelization is developed for CMB maps.
The scheme is based on the Gauss--Legendre polynomials zeros
and allows one to create strict orthogonal expansion of the map.
A corresponding code has been implemented and comparison with other
methods has been done.

}
\maketitle
\markboth
{Doroshkevich et~al.: Gauss--Legendre sky pixelization for CMB maps}
{Doroshkevich et~al.: Gauss--Legendre sky pixelization for CMB maps}

\section{Introduction}
Starting from the COBE experiment using the so called Quadrilateralized
Sky Cube Projection (see Chan and O'Neill$^1$ 1976, O'Neill and Laubscher$^2$
1976, Greisen and Calabretta$^3$ 1993), the problem of the whole sky CMB
pixelization has attracted great interest.
At least three methods of the CMB
celestial sphere pixelization have been proposed and implemented after
the COBE pixelization scheme:
the Icosahedron pixelizing by Tegmark$^4$ (1996),
the Igloo pixelization by Crittenden and Turok$^5$ (1998, hereafter CT98)
and the HEALPix\footnote{currently http://www.eso.org/science/healpix/}
  method by G\'orski et al.$^6$ (1999) (with the last modification in 2003).
Two important questions mentioned already by Tegmark$^4$ (1996)
are now under discussion: a) what is the optimal method for the choice
of the $N_{pix}$ positions of pixel centers, shapes and
sizes to provide (as good as possible) the compact uniform coverage
of the sky by pixels with equal areas, and b) what is the best
way to approximate any convolutions of the maps by sums using pixels ?

All the above mentioned pixelization schemes were devoted to solve
the first problem as accurate as possible, and the answer to the  
second question usually follows for the chosen pixelization scheme.

In this paper we change the focus of the problem to processing
on the sphere and then determine the scheme of pixelization. We would
like to remind that pixelization of the CMB data on the sphere is 
only some part of the general problem, which is the determination of
the coefficients of the spherical harmonic decomposition of the 
CMB signal for both anisotropy and polarization.
These coefficients, which we call $a_{\ell m}$, are used in subsequent steps
in the analysis of the measured signal, and in particular, in the
determination of the power spectra, $C_{\ell}$, of the anisotropy and
polarization (see review in Hivon at al,$^{7}$ 2002), in some special
methods for components separation (Stolyarov et al.$^{8}$ 2002; Naselsky et
al.$^{9}$ 2003a) and phase statistics (Chiang et al.$^{10}$ 2003; Naselsky et
al.$^{11,12}$ 2003b, 2004; Coles et al.$^{13}$ 2004).

Here we propose a specific method to calculate the coefficients
$a_{\ell m}$. It is based on the so called Gaussian quadratures and is
presented in Sec. 2\,. In this specific pixelization scheme correspond
the position of pixel centers along the $\theta$--coordinate to
so--called the Gauss--Legendre quadrature zeros and it will be shown
(Sec. 5) that this method increases the accuracy of calculations essentially.

Thus, the method of calculation of the coefficients $a_{\ell m}$
dictates the method of the pixelization. We call our method GLESP, the
Gauss--Legendre Sky Pixelization. We have developed a special code for
the GLESP approach and a package of codes which are necessary for the
whole investigation of the CMB data including the determination of
anisotropy and polarization power spectra, $C_\ell$, the Minkowski
functionals and other statistics. 

This paper is devoted to description of the main idea of the GLESP
method, the estimation of the accuracy of the different steps and
of the final results,  the description of the GLESP code and
its testing.
We do not discuss the problem of
integration over a finite pixel size for the time ordered data
in this paper.
The simplest scheme of integration over pixel area
is to use equivalent
weight relatively to the center of the pixel. The GLESP code
uses this method as HEALPix and Igloo do.

In forthcoming papers we shall
discuss our GLESP code extension
on the processing of CMB polarization data,
the Minkowski functionals and
the peak statistics of  CMB maps.

\section{Main ideas and basic relations}

The standard decomposition of the measured temperature variations 
on the sky, $\Delta T(\theta,\phi)$, in spherical harmonics is
\begin{equation}
\Delta T(\theta,\phi)= \sum_{\ell=2}^{\infty}\sum_{m=-\ell}^{m=\ell} a_{\ell m}
Y_{\ell m} (\theta, \phi)\
\label{eq1}                  
\end{equation}
\be
 Y_{\ell m}(\theta,\phi) = \sqrt{{(2\ell+1)\over 4\pi}{(\ell-m)!
\over (\ell+m)!}}P_\ell^m(x) e^{i m\phi},~x=\cos\theta\,,
\label{eq2}                  
\ee
where $P_\ell^m(x)$ are the associated Legendre polynomials.
For a continuous $\Delta T(x,\phi)$ function, the coefficients 
of decomposition, $a_{\ell m}$, are
\begin{equation}
a_{\ell m}=\int^1_{-1}dx\int^{2\pi}_0 d\phi\Delta T(x,\phi)
Y^{*}_{\ell m}(x,\phi)
\label{eq3}                  
\end{equation}
where $Y^{*}_{\ell m}$ denotes complex conjugation of $Y_{\ell m}$.
For numerical evaluation of the integral Eq(\ref{eq3}) we will use the
Gaussian quadratures, a method which was proposed by Gauss in 1814,
and developed later by Christoffel in 1877.
As the integral over $x$ in  Eq(\ref{eq3}) is an integral over a polynomial
of $x$ we may use the
following equality (Press et al.$^{14}$ 1992)
\be
\int^1_{-1}dx \Delta T(x,\phi) Y^{*}_{\ell m}(x,\phi)=\sum^N_{j=1}
w_j\Delta T(x_j,\phi) Y^{*}_{\ell m}(x_j,\phi)\,.
\label{eq4}
\ee
where $w_j$ is a proper Gaussian quadrature weighting function.
Here the weighting function $w_j=w(x_j)$ and $\Delta T(x_j,\phi)
Y^{*}_{\ell m}(x_j,\phi)$ are taken at points $x_j$ which are the
net of roots of the Legendre polynomial
\be
P_N(x_j)=0\,,
\label{root}
\ee
where $N$ is the 
maximal rank of the polynomial under consideration. It is well known 
that the equation $P_N(x_j)=0$ has $N$ number of zeros in
interval $-1\le x\le 1$. For the Gaussian--Legendre method Eq(\ref{eq4}),
the weighting coefficients are (Press et al.$^{14}$ 1992)
\be
w_j= {2\over 1-x^2_j} [P_N^{'}(x_j)]^{-2}\,,
\label{eq6}
\ee
where ${'}$ denotes a derivative. They can be calculated together
with the set of $x_j$
with the `{\tt gauleg}' code (Press et al.$^{14}$ 1992, Sec. 4.5).

\begin{figure}[!th]
\psfig{figure=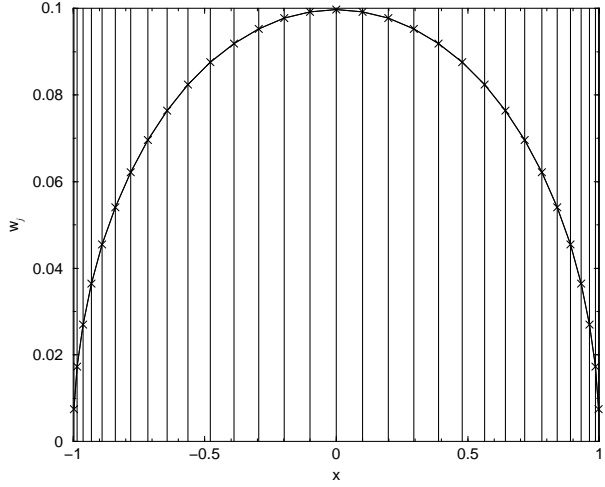,width=8cm,angle=-90}
\caption{Gauss-Legendre weighting coefficients ($w_j$) versus
Legendre polynomial zeros ($x_j=\cos\theta_j$)
being centers of rings used in GLESP for the case
of $N=31$. Positions of zeros are plotted by vertical lines.
}
\label{fig_weights}
\end{figure}

\begin{figure}[!th]
\psfig{figure=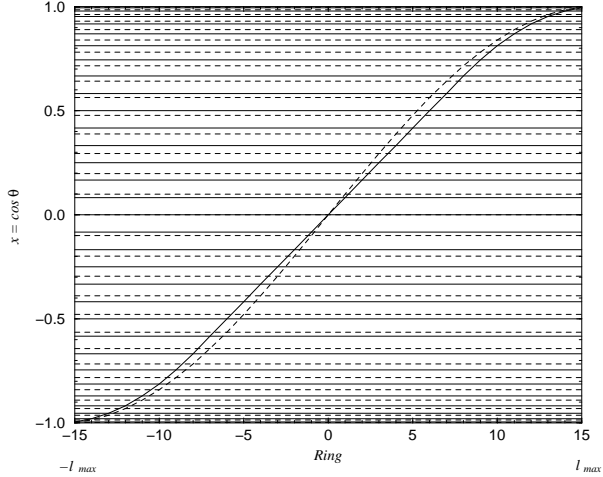,width=8cm,angle=-90}
\caption{Ring center position ($x_j=\cos \theta_j$) vs ring number
for 2 pixelization schemes,
HEALPix (solid) and GLESP (dashed).
Figure demonstrates the case of $N=31$.
}
\label{fig_cos_theta}
\end{figure}

\begin{figure*}[!th]
\centerline{
\vbox{
\psfig{figure=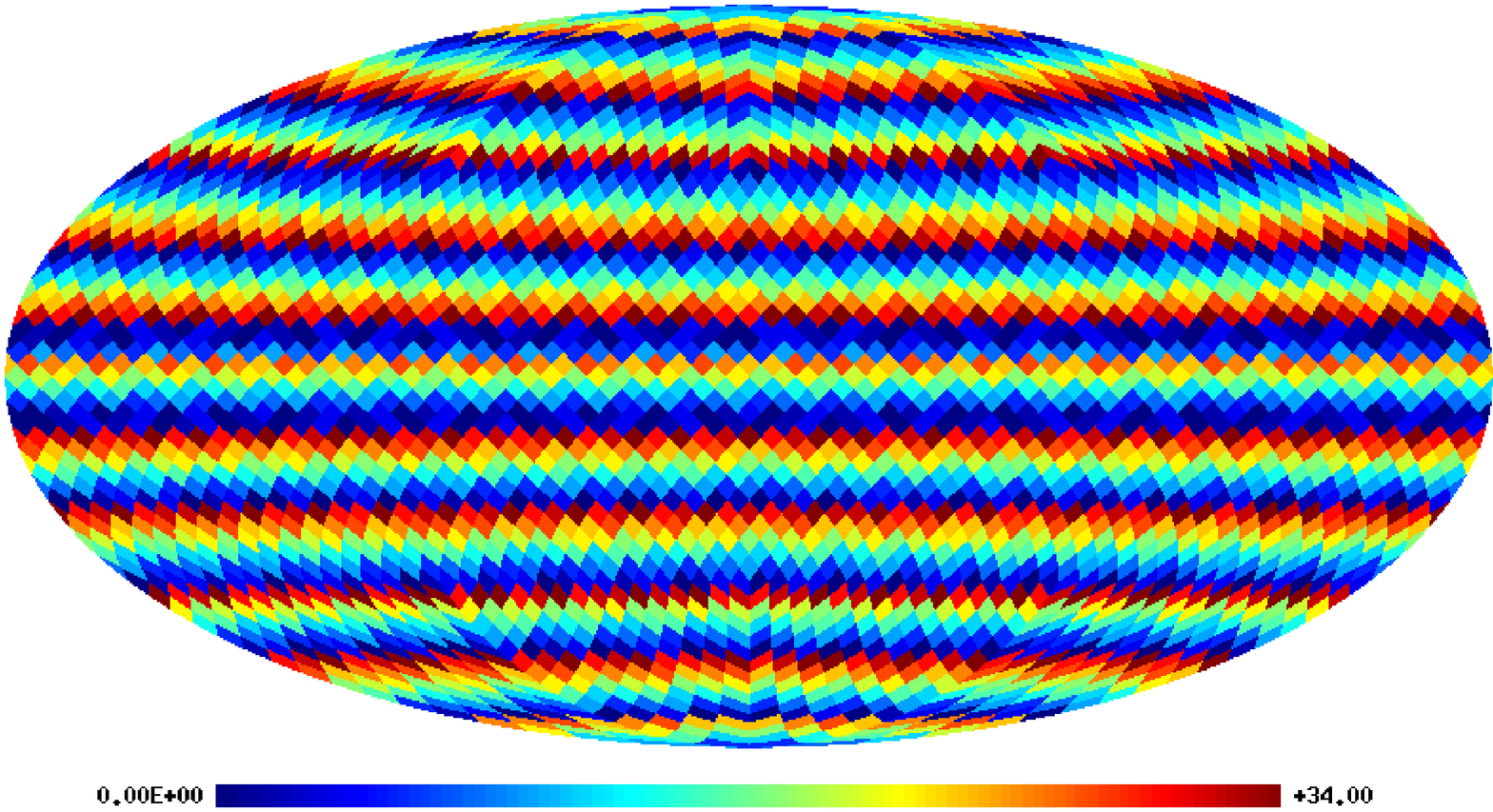,width=12cm,angle=-90,bbllx=90pt,bblly=40pt,bburx=475pt,bbury=800pt,clip=0}
\psfig{figure=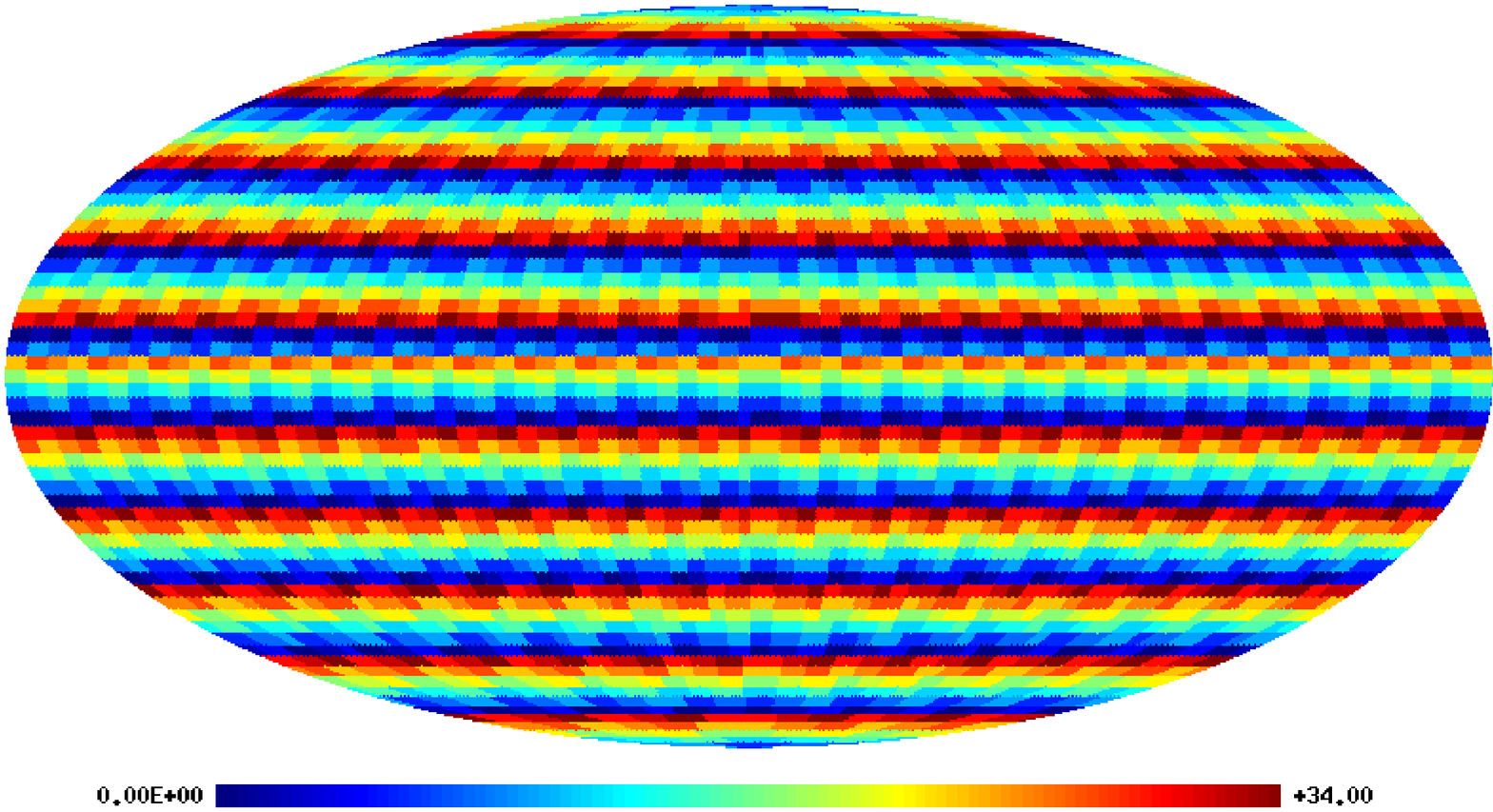,width=12cm,angle=-90,bbllx=90pt,bblly=40pt,bburx=475pt,bbury=800pt,clip=0}
}
}
\caption{Schematic representation of
2 types of pixelization on sphere:
HEALPix (top) and GLESP (bottom).
Various color of pixels is used to show their shape.
}
\label{fig_pix_map}
\end{figure*}

In the GLESP approach are the trapezoidal pixels
bordered by $\theta$ and $\phi$ coordinate lines with the pixel centers 
(in the $\theta$ direction) situated at points with $x_j=\cos\theta_j$.
Thus, the interval
$-1\leq x\leq 1$
is covered by $N$ rings of the pixels (details are given in Sec. 3).
The angular resolution achieved in the measurement of the CMB data determines
the upper limit of summation in Eq. (\ref{eq1}), $\ell\leq \ell_{max}$. To
avoid the Nyquist restrictions we use a number of pixel rings,
$N\geq 2 \ell_{max}$.
In order to make the pixels in the equatorial ring (along the $\phi$
coordinate) nearly squared, the number of pixels in this direction
should be $N_\phi^{max}\approx 2N$.
The number of pixels in other rings, $N_\phi^j$,
must be determined from the condition of
making the pixel sizes as equal as possible with the equatorial ring of pixels.

Fig. \ref{fig_weights} shows the weighting coefficients,
$w_j$, and the position of pixel centers
for the case $N=31$.
Fig. \ref{fig_cos_theta} compares some features of the pixelization schemes
used in HEALPix and GLESP
(see Sec.\,4).
Fig. \ref{fig_pix_map} compares pixel distribution and shapes on a sphere in
the mollview projections of HEALPix and in GLESP.

In the definition (\ref{eq1}) are the coefficients $a_{\ell m}$
complex quantities while $\Delta T$ is real.
In the GLESP code started from the definition (\ref{eq1}) we use
the following representation of the $\Delta T$ 
\begin{eqnarray}
\Delta T(\theta,\phi)& =& \sum_{\ell=2}^{\ell_{max}} a_{\ell0} Y_{\ell0}(\theta,\phi) \nonumber \\
      & + &   \sum_{\ell=2}^{\ell_{max}}\sum_{m=1}^\ell\left[a_{\ell m}
          Y_{\ell m}(\theta,\phi)+a_{\ell,-m}Y_{\ell,-m}(\theta,\phi)\right],
\label{eq7}
\end{eqnarray}
where
\begin{equation}
Y_{\ell,-m}(\theta,\phi)=(-1)^mY^*_{\ell,m}(\theta,\phi),\quad 
a_{\ell m}=(-1)^m a^{*}_{\ell,-m}\,.
\label{eq8}
\end{equation}
Thus,
\begin{eqnarray}
\Delta T(\theta, \phi)&=&
\frac{1}{\sqrt{2\pi}}\sum_{\ell=2}^{\ell_{max}} \sqrt{\frac{2 \ell+1}{2}}Re(a_{\ell,0})P_\ell^0(\cos\theta) \nonumber \\
&+&\sqrt{\frac{2}{\pi}}\sum_{\ell=2}^{\ell_{max}}\sum_{m=1}^\ell
\sqrt {\frac{2\ell+1}{2}\frac{(\ell-m)!}{(\ell+m)!}}P_\ell^m(\cos\theta) \times \nonumber \\
&& \left[Re(a_{\ell m})\cos(m\phi)- Im(a_{\ell m})\sin(m\phi)\right]
\label{eq9}
\end{eqnarray}
where $P_\ell^m(\cos\theta)$ are the well known associated Legendre polynomials
(see Gradshteyn and Ryzhik$^{15}$ 2000).
In the GLESP code, we use
normalized associated Legendre polynomials $f^m_\ell$:
\be
f^m_\ell(x)=\sqrt {\frac{2\ell+1}{2}\frac{(\ell-m)!}{(\ell+m)!}}P^m_\ell(x)
\label{eq10}
\ee
where $x=\cos\theta$, and $\theta$ is the polar angle.
These polynomials, $f_\ell^m(x)$, can be calculated using two well known
recurrence relations. The first of them gives $f_\ell^m(x)$ for a
given $m$ and all $\ell>m$:
\begin{equation}
f_\ell^m(x) = x\sqrt{4\ell^2-1\over \ell^2-m^2}f_{\ell-1}^m -\sqrt{{2\ell+1\over 2\ell-3}
     {(\ell-1)^2-m^2\over \ell^2-m^2}}
f_{\ell-2}^m
\label{legm}
\end{equation}
This relation starts with
\[
f _m^m(x)={(-1)^m\over\sqrt{2}}\sqrt{(2m+1)!!\over{(2m-1)!!}}
(1-x^2)^{m/2},\quad
\]
\[
f_{m+1}^m=x\sqrt{2m+3}f_m^m
\]
The second recurrence relation gives $f_\ell^m(x)$ for a given $\ell$ and
all $m\leq l$:   
\[
\sqrt{(\ell-m-1)(\ell+m+2)}f_\ell^{m+2}(x)+{2x(m+1)\over\sqrt{1-x^2}}
f_\ell^{m+1}(x)+
\]
\be
\sqrt{(\ell-m)(\ell+m+1)}f_\ell^m(x)=0\,,
\label{legl}
\ee
This relation is started with the same $f_\ell^\ell(x)$ and $f_\ell^0(x)$
which must be found with (\ref{legm}). 

As discussed in Press et al.$^{14}$ (1992, Sec. 5.5), the first 
recurrence relation (\ref{legm}) is formally unstable if the number 
of iteration is going to infinity. Unfortunately, there are no 
theoretical recommendations what the maximum iteration one can 
use in the quasi-stability area. However, it can be used because
we are interested in the so called {\it dominant} solution 
(Press et al.$^{14}$ 1992, Sec. 5.5), which is approximately stable.
The second recurrence relation (\ref{legl}) is stable for all 
$\ell$ and $m$.

\section{Properties of GLESP}

Following the previous discussion
we define the new pixelization scheme GLESP as follows:
\begin{itemize}
\item 
    In the polar direction $x=\cos\theta$, we define
    $x_j, j=1,2,\ldots,N$, as
    the net of roots
    of Eq. (\ref{root}).
\item 
    Each root $x_j$ determines the position of a ring with
    $N_{\phi}^j$ pixel centers with $\phi$--coordinates $\phi_i$.
\item
    All the pixels have nearly equal area.
\item
    Each pixel has weight $w_j$
        (see Eq (\ref{eq6})).
\end{itemize}

\begin{figure}[!h]
\vbox{
\psfig{figure=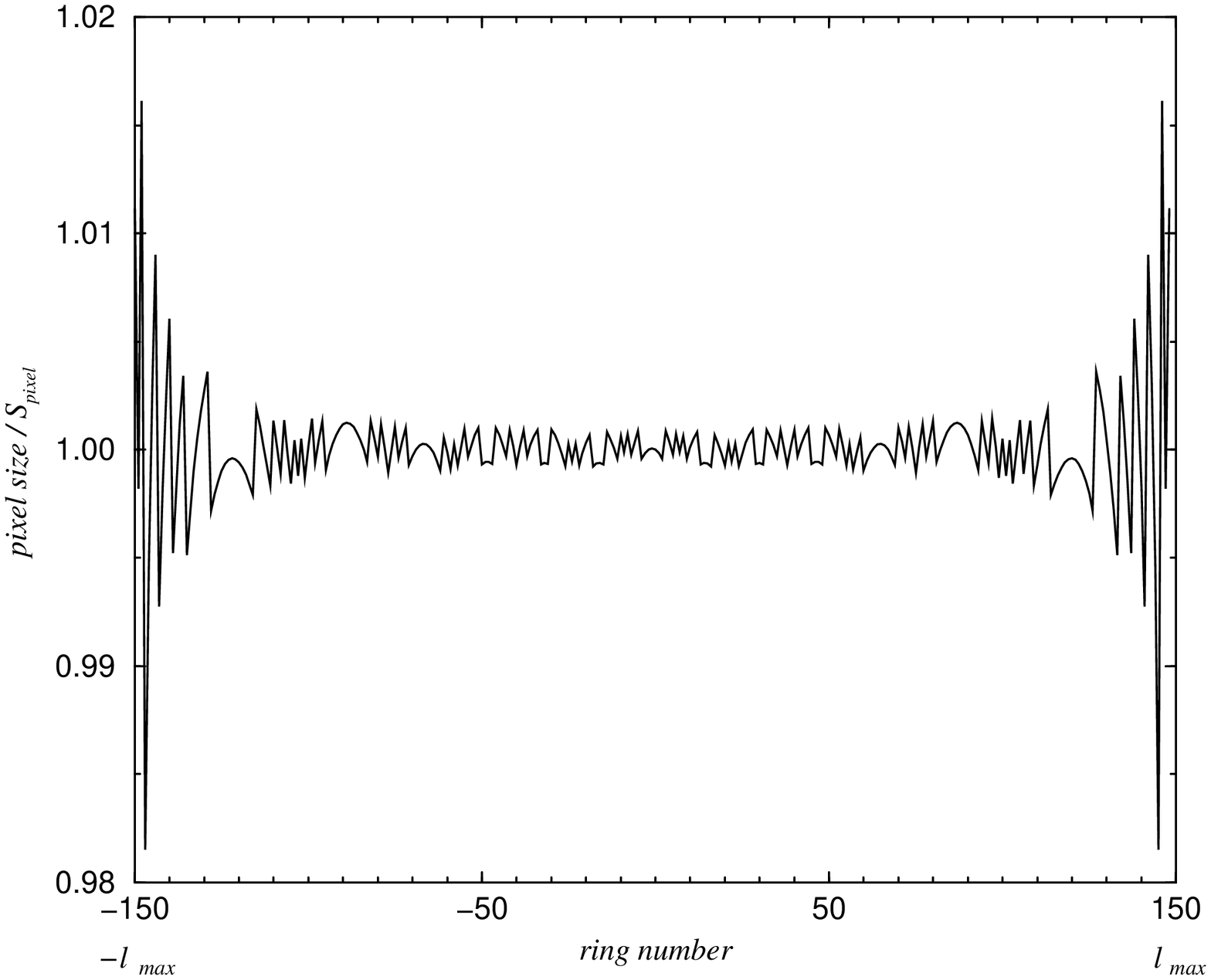,width=8cm,angle=-90}
\psfig{figure=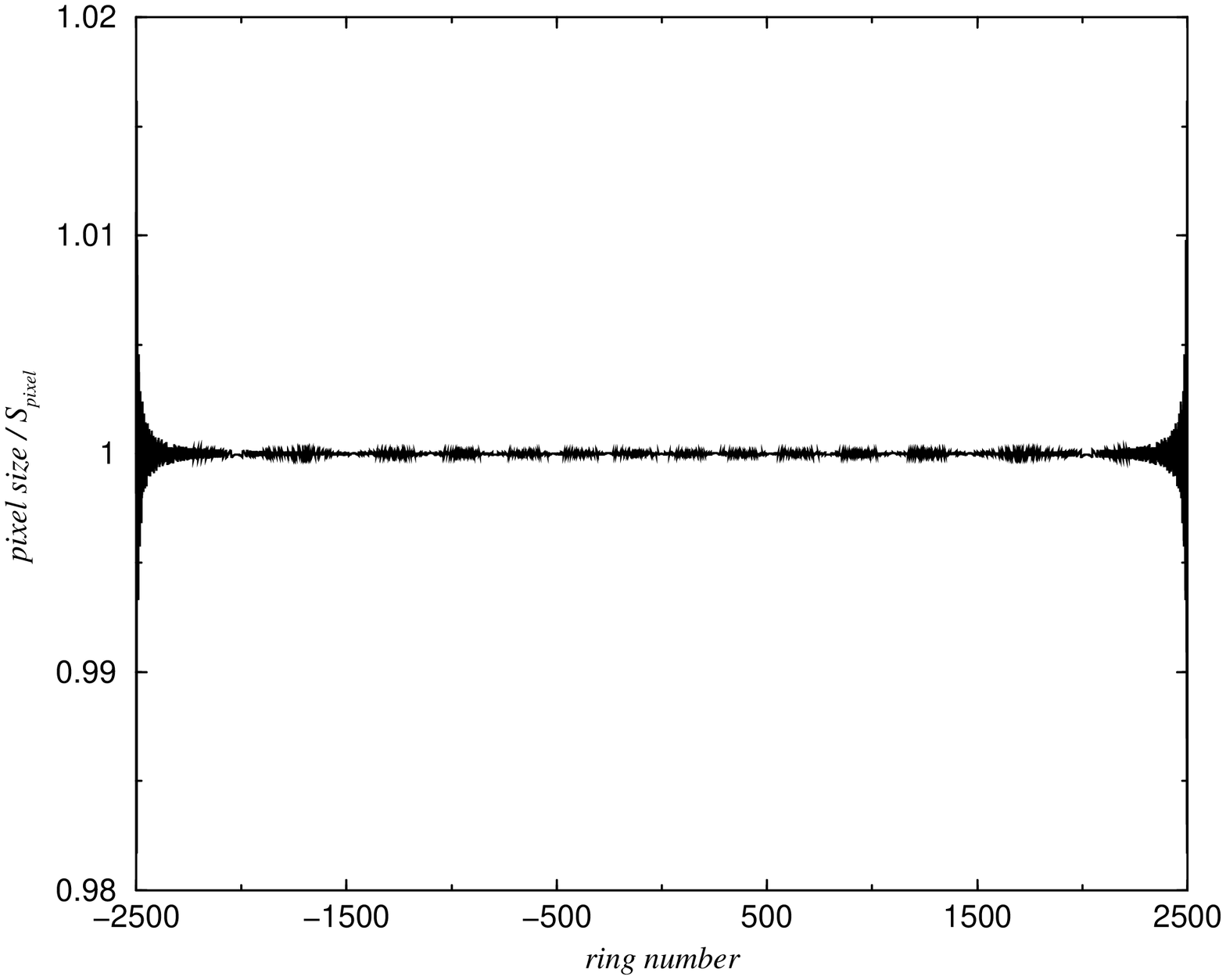,width=8cm,angle=-90}
}
\caption{
Pixel size/equator pixel area vs ring-number
for GLESP for number of rings $N=300$ and $N=5000$.}
\label{fig_pix_size}
\end{figure}

In our numerical code which realizes the GLESP pixelization scheme we 
use the following conditions.
\begin{itemize}
\item
       Borders of all pixels are along the coordinate lines of $\theta$
       and $\phi$. Thus with a reasonable accuracy they are trapezoidal.
\item
       The number of pixels along the azimuthal direction $\phi$ depends
       on the ring number. The code allows to choose an arbitrary
       number of these pixels. Number of pixels depends on the 
       $\ell_{max}$ accepted for the CMB data reduction.
\item
       To satisfy the Nyquist's theorem, the number $N$ of the ring
      along the $x=\cos \theta $ axis must be taken as $N\geq 2 \ell_{max}+1$.
\item
       To make equatorial pixels roughly square, the number
       of pixels along the azimuthal axis, $\phi$, is taken as
       $N^{max}_\phi = {\rm int}(2\pi/d\theta_k +0.5)$, where
       $k={\rm int}(N+1)/2$, and $d\theta_k = 0.5(\theta_{k+1}-
       \theta_{k-1})$.
\item
       The nominal size of each pixel is defined as
       $S_{pixel}=d\theta_k{\times}d\phi$,
       where
       $d\theta_k$ is the value on the equatorial ring and
       $d\phi = 2\pi / N^{mx}_\phi$ on equator.
\item
       The number $N_\phi^j$ of pixels in the $j^{th}$ ring at $x=x_j$ is
        calculated
       as $N_\phi^j = {\rm int}(2\pi \sqrt{1-x_j^2}/S_{pixel}+0.5)$;
\item
       Polar pixels are triangular.
\item
       Because the number $N_\phi^j$ differs from $2^k$ where $k$ is
        integer, we use for the Fast Fourier transform along the
        azimuthal direction the FFTW code (Frigo and Johnson$^{16}$ 1997).
        This code permits one to use not only 2$^n$--approach, but
        other base-numbers too, and provide even faster speed.

\end{itemize}

\begin{figure}[!h]
\psfig{figure=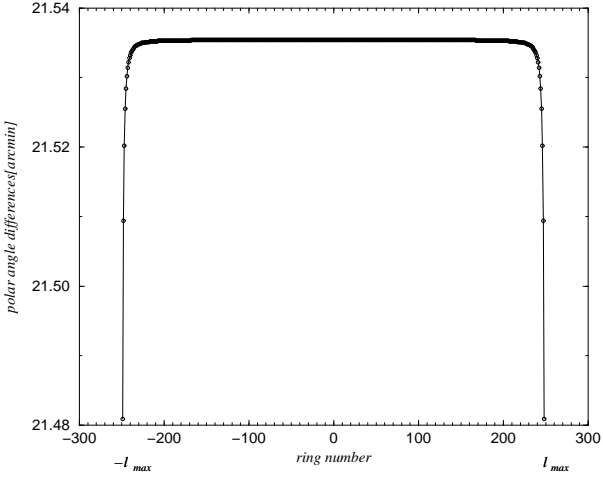,width=8cm,angle=-90}
\caption{Pixel size along polar angle ($\ell_{max}=250$)}
\label{fig_layer_dist}
\end{figure}

With this scheme,
the pixel sizes  are equal inside each ring, and
with a maximum deviation between the different rings of
$\sim$1.5\% close to the poles
(Fig\,\ref{fig_pix_size}).
Increasing resolution decreases an absolute error of an area
due to the in-equivalence of polar and equator pixels proportionally to
$N^{-2}$.

Fig. \ref{fig_layer_dist} shows that this pixelization scheme
for high resolution maps (e.g. $\ell_{max}>500$) produces nearly
equal thickness $d\theta$ for the most rings.

GLESP has not the hierarchical structure, but the problem
of the closest pixel selection is on the software level.
Despite GLESP is close to the Igloo pixelization scheme in the azimuthal
approach, there is a difference between the two schemes
in connection with
the $\theta$-angle (latitude) pixel step selection.
Therefore, we can not unify these two
pixelizations. The Igloo scheme applied to the GLESP latitude step will
give too different pixel areas. The pixels will be neither equally
spaced in
latitude, nor  of uniform area, like Igloo requires.

\section{GLESP pixel window function}

For application of the GLESP scheme,
we have to take into account the influence of the pixel size,  
shape and its location on the sphere on the signal in the pixel
and its contribution to the power spectrum $C(\ell)$. The temperature
in a pixel is (G\'orski et al.$^{6}$ 1998; CT98)
\begin{equation}
\Delta T_p=\int_{\Delta\Omega_p} W_p(\theta, \phi)\Delta 
T(\theta,\phi)d\Omega
\label{eq13}
\end{equation}
where $W_p(\theta,\phi)$ is the window function
for the $p$-th pixel with the area $\Delta\Omega_p$.
For the window function $W_p(\theta,\phi)=1$ inside the
pixel and $W_p(\theta,\phi)=0$ outside (G\'orski et al.$^{6}$ 1998),
we have from Eq(\ref{eq1}) and Eq(\ref{eq13}): 
\[
\Delta T_p=\sum_{\ell,m}a_{\ell m}W_p(\ell,m)
\]
where
\[
W_p(\ell,m)= \int d\Omega W_p( \theta, \phi)Y_{\ell m}(\theta,\phi)
\]
and
\begin{eqnarray}
 W_p( \theta, \phi )= \sum_{\ell,m} W_p(\ell, m)Y^{*}_{\ell m}( \theta, \phi).
\label{eq14}
\end{eqnarray}
\noindent
The corresponding correlation function (CT98) for the pixelized
signal is

\begin{equation}
\langle \Delta T_p\Delta T_q \rangle= \sum_{l,m} C(\ell) W_p(\ell,m)
W^{*}_q(\ell,m)
\label{eq15}
\end{equation}

\subsection{Accuracy of the window function estimation}

The discreetness of the pixelized map determines the properties of
the signal for any pixels and restricts the precision achieved 
in any pixelization scheme. To estimate this precision we can 
use the expansion (CT98)
\begin{equation}
\Delta T^{map}( \theta,\phi ) = \sum_p S_p\Delta T_p W_p(\theta,\phi) =
\sum_{\ell,m}a^{map}_{\ell m}Y_{\ell m}(\theta,\phi)
\label{eq16}
\end{equation}
\begin{eqnarray}
a^{map}_{\ell m}& =& \int d\Omega\Delta T^{map}(\theta,\phi)
      Y^{*}_{\ell m}(\theta,\phi) \nonumber \\
& = & \sum_p S_p\Delta T_p W^{*}_p(\ell,m),  \label{eq17} 
\end{eqnarray}
where $S_p$ is the area of the $p$-th pixel.
These relations generalize
Eq. (\ref{eq2}), taking properties of the window function into account.
The GLESP scheme uses the properties of Gauss--Legendre
integration in the polar direction while azimuthal pixelization for each 
ring is similar to the Igloo scheme and we get (see Eq(4)).
\begin{eqnarray}
W_p(\ell,m)& =&  \frac{w_p}{\sqrt{2\pi}\Delta x_p}\exp\left(\frac{im\pi}
    {N_\phi^p}\right) \nonumber \\
    & \times &\frac{\sin\left(\pi m/N_\phi^p\right)}
    {\left(\pi m/N_\phi^p\right)} \int_{x_p-0.5\Delta x_p}^{x_p+0.5\Delta x_p} dxf_\ell^m(x)
\label{eq18}
\end{eqnarray}
where $\Delta x_p=(x_{p+1}-x_{p-1})/2$ with $x_p$ the p-th Gauss--Legendre
knot and $N_\phi^p$ the number of pixels in the azimuthal direction.
This integral can be rewritten as follows:
\[
\int_{x_p-0.5\Delta x_p}^{x_p+0.5\Delta x_p} dx f_\ell^m(x)
   \simeq
\]
\begin{equation}
\sum_{k=0}^{\infty}\frac{1+(-1)^{k}}{(k+1)!}{f^{(k)}}_\ell^m(x_p)
\left(\frac{\Delta x_p}{2}\right)^{k+1}
\label{eq19}
\end{equation}
where ${f^{(k)}}_\ell^m(x_p)$ denotes the k-th derivatives at
$x=x_p$. So, for $\Delta x_p\ll 1$ we get the expansion of (\ref{eq18}):
\begin{equation}
W_p^{(2)}(\ell,m) = W_p^{(0)} (\ell,m) \left(1+\frac{f^{(2)m}_\ell(\Delta x_p)^2}
             {24f_\ell^m}\right)
\label{eq20}
\end{equation}
where
\begin{equation}
W_p^{(0)}(\ell,m) \simeq \frac{w_p}{\sqrt{2\pi}}\exp\left(\frac{im\pi}
    {N_\phi^p}\right)\frac{\sin\left(\pi m/N_\phi^p\right)}
    {\left(\pi m/N_\phi^p\right)}f_\ell^m(x_p)
\label{eq21}
\end{equation}
where
is independent of $\Delta x_p$.
For the accuracy of this estimate we get
\begin{equation}
\frac{\delta W_p(\ell,m)}{W_p (\ell,m)} =
    \frac{W_p^{(2)}(\ell,m)- W_p^{(0)}(\ell,m)}{W_p^{(0)}(\ell,m)} \simeq
      \left|\frac{(f^{\prime\prime})_\ell^m (\Delta x_p)^2}{24 f_\ell^m} \right|
\label{eq22}
\end{equation}

According to the last modification of the HEALPix, an accuracy of 
the window function reproduction is about $10^{-3}$.
To obtain
the same accuracy for the $W_p(\ell,m)$, we need to have
\begin{equation}
\Delta x_p\le 0.15 \left.\left|\frac{f_\ell^m}{(f^{\prime\prime})_\ell^m}
               \right|^{\frac{1}{2}}\right|_{x=x_p}
\label{eq23}
\end{equation}
Using the approximate link between Legendre and Bessel functions
for large $\ell$ (Gradshteyn and Ryzhik$^{15}$ 2000)
\[
f_\ell^m\propto J_m(\ell x)
\]
  we get:
\begin{equation}
\Delta x_p\le 0.15 x_p/\sqrt{m(m+1)} 
\label{eq24}
\end{equation}
and for $\Delta x_p\sim \pi/N$ we have from Eq.(\ref{eq24})

\begin{equation}
\frac{\delta W_p(\ell,m)}{W_p(\ell,m)}\geq 10^{-2}\cdot\left({\ell_{max}\over
    N}\right)^2
\label{eq25}
\end{equation}
For example, for $N=2 \ell_{max}$, we obtain $\delta W_p(\ell,m) /
 W_p(\ell,m)\simeq 2.3\times10^{-3}$, what is a quite reasonable accuracy
for $\ell_{max}\sim$ 3000--6000.

\begin{figure*}[!th]
\centerline{
\psfig{figure=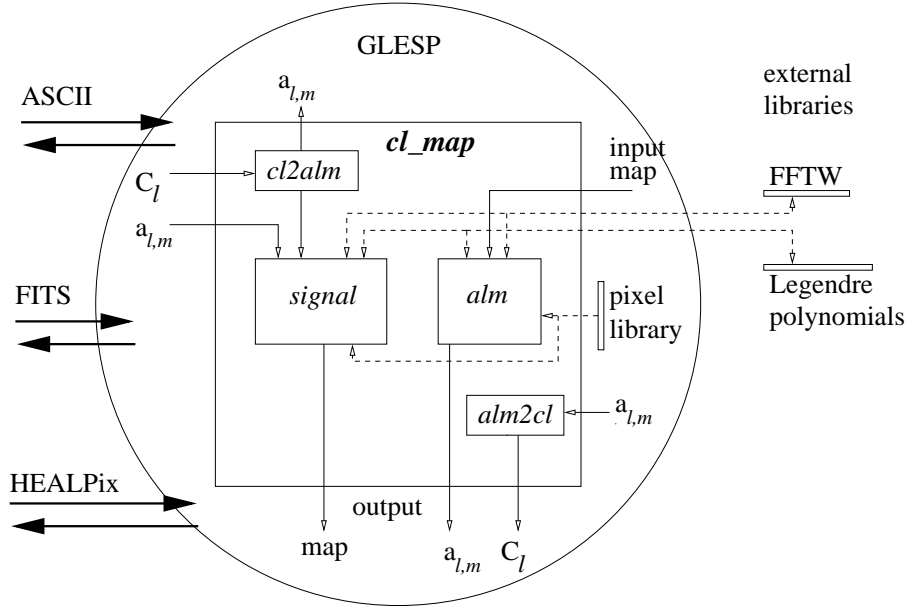,width=12cm,angle=-90}
}
\caption{
Structure of the GLESP package.
}
\label{glesp_scheme}
\end{figure*}

\section{Structure of the GLESP  code}

The code is developed in two levels of organization.
The first one, which unifies F77 FORTRAN and C functions,
subroutines
and wrappers for C routines to be used
for FORTRAN calls, consists of the main procedures '{\tt signal}' which
transforms given values of $a_{\ell m}$ to a map, '{\tt alm}' which
transforms a map to $a_{\ell m}$, '{\tt cl2alm}' which creates a sample of
$a_{\ell m}$ coefficients for a given $C_\ell$ and '{\tt alm2cl}' which
calculates $C_\ell$ for $a_{\ell m}$.
Procedures for code testing, parameters control
Kolmogorov-Smirnov analysis for Gaussianity of $a_{\ell m}$
and homogeneity of phase distribution,
and others, are also included.
Operation of these routines is based on a block of procedures
calculating the Gauss--Legendre pixelization for a given resolution
parameter, transformation of angles to pixel numbers and back.

The second level of the package contains the programs
which are convenient for the utilization of the first level routines.
In addition to the straight use of the already mentioned four main
procedures, they also provide means to calculate map patterns
generated by the $Y_{20}$, $Y_{21}$ and
$Y_{22}$ spherical functions,
to compare two sets of $a_{\ell m}$ coefficients,
to convert a GLESP map to a HEALPix map,
to convert a HEALPix map, or other maps, to a GLESP map

Fig. \ref{glesp_scheme} outlines the GLESP package.
The circle defines the zone of the GLESP influence based on the pixelization
library. It can include several subroutines and operating programs.
The basic program '{\tt cl\_map}' of the second level,
shown as a big rectangle, interacts
with the first level subroutines.
These subroutines are shown by small rectangles and call
external libraries for the Fourier transform and
Legendre polynomial calculations.
The package reads and writes data
both in ASCII table and FITS formats.
More than 10 programs of the GLESP package operate in
the GLESP zone.

The present development of the package has also
parallel calculation
implementation.
Visualization procedures in OPEN\,GL have been developed at IaO, Cambridge.

\subsection{Test and precision of the GLESP code}

Three tests allow us to check the code. The first of them is
from the analytical maps
\[
Y_{2,0}=\sqrt{5\over 16\pi}(3x^2-1)\,,
\]
\[
 Y_{2,1}=-\sqrt{15\over 8\pi}x\sqrt{1-x^2}\cos\phi,\quad 
\]
\[
Y_{2,-1}=-\sqrt{15\over 8\pi}x\sqrt{1-x^2}\sin\phi\,,
\]
\[
 Y_{2,2}=\sqrt{15\over 32\pi}(1-x^2)\cos(2\phi),\quad 
\]
\[
Y_{2,-2}=-\sqrt{15\over 32\pi}(1-x^2)\sin(2\phi)\,.
\]
to calculate $a_{\ell m}$. The code reproduces the theoretical $a_{\ell m}$
better than $10^{-7}$.

The second test is to reproduce an analytical map
$\Delta T(x,\phi)=Y_{\ell m}(x,\phi)$ from a given $a_{lm}$
These tests check the calculations of the
map and spherical coefficients independently.

The third test is the reconstruction of
$a_{\ell m}$ after the
calculations of the map, $\Delta T(x,\phi)$, and back.
This test allows one to check orthogonality.
If the transformation is based on really orthogonal functions
it has to return after forward and backward calculation the
same $a_{\ell m}$  values.

Precision of the code can be estimated by introduction
of a set of $a_{\ell m}=1$ and reconstruction of them. This
test showed that using relation (\ref{legm}) we can
reconstruct the introduced $a_{\ell m}$ with the precision $\sim
10^{-7}$ limited only by single precision of float point data recording
and with the precision $\sim 10^{-5}$ for relation
(\ref{legl}).

\begin{figure}[!h]
\centerline{
\psfig{figure=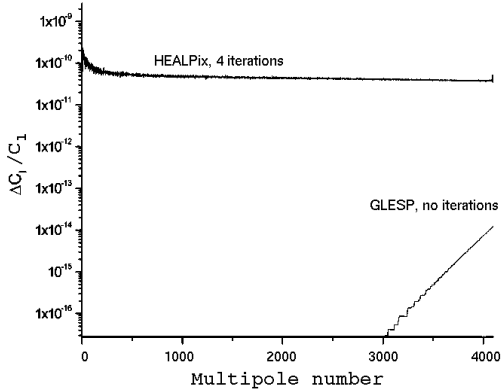,width=8cm,angle=-90}
}
\caption{
Comparison of calculation accuracy in HEALPix (4 iterations) of
the version 1.20 and
in GLESP (no iterations)  
methods.
Number of pixels is approximately the same ($\sim6\times10^7$)
and calculation time is proportional to the number of
iterations. Note that the comparison is based on iterations of {\it
constant} signal. For real signal (where $a_{\ell m} \ne$ constant), the
accuracy is proportional to the square root of those shown in the figure,
i.e. $ \Delta C_\ell /C_\ell < 10^{-8}$ for GLESP at $\ell
< 3000$ whereas $\sim 10^{-5}$ for HEALPix at all multipoles.}   
\label{hp_compare}
\end{figure}

Fig. \ref{hp_compare} demonstrates the accuracy of $C_\ell$ calculations
using HEALPix and GLESP\footnote{Calculations were
carried out and Fig.\ref{hp_compare}
was produced by Vlad Stolyarov at IoA, Cambridge}.

It should be noted that unlike the HEALPix code, the GLESP
method does not needed any iteration for calculation of
the $a_{\ell m}$ coefficients
and therefore is much faster.
Our definition of the $a_{\ell m}$ coefficients is exactly
the same as in HEALPix as an estimator of the anisotropy power spectrum:
\begin{equation}
C(\ell)=\frac{1}{2\ell+1}\left[|a_{\ell 0}|^2 + 2\sum_{m=1}^\ell |a_{\ell m}|^2\right]
\label{eq32}
\end{equation}

\subsection{Re-pixelization}

Any re-pixelization procedure will cause loss of information and
thereby introduce uncertainties and errors.
The GLESP code has procedures for map re-pixelization based
on two different methods in the $\Delta T(\theta,\phi)$--domain:
the first one consists in
averaging input values in the corresponding pixel,
the second one is connected
with spline interpolation inside the pixel grid.

In the first method, we consider input pixels which fell in our pixel
with values
$\Delta T(\theta_{i},\phi_{i})$
to be averaged with
a weighting function. The realized weighting function is
a function of simple averaging with equal weights.
This method is widely used in appropriation of a given values to
the corresponding pixel number.

\begin{figure}[!h]
\psfig{figure=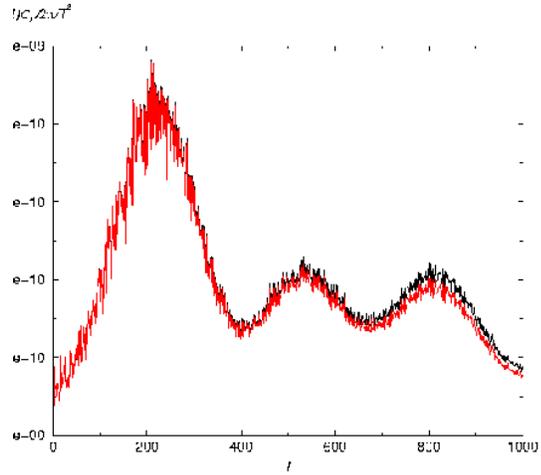,width=8cm,angle=0,bbllx=55pt,bblly=45pt,bburx=555pt,bbury=640pt,clip=}
\caption{Power spectra calculated for the initial HEALPix map (black curve)
with
$\ell_{max}=1000$,
$N_{side}=1024$, pixel size = \mbox{11.8026$\sq^{\prime}$},
and $N_{tot}$=12\,582\,912, and for resulting re-pixelized GLESP map
(red curve)
with the closest possible pixel size = \mbox{11.8038$\sq^{\prime}$},
$N_{tot}$=12\,581\,579.
Deviations of the power spectra at high $\ell$ illustrate the ratio of the
HEALPix and GLESP window functions.
}
\label{fig_hpgl_rep}
\end{figure}

In the second method of re-pixelization, we use a spline interpolation
approach. If we have a map $\Delta T(\theta_i,\phi_i)$ recorded in 
the knots different from the Gauss--Legendre grid, it is possible to 
repixelize it to our grid $\Delta T(\theta_{i}^{\prime},\phi_i^{\prime})$
using approximately the same number of pixels
and the standard interpolation scheme based on the cubic spline 
approach for the map re-pixelization.
This approach is sufficiently fast because
the spline is calculated once for one vector of the tabulated data
(e.g. in one ring), and values of interpolated function for any
input argument are obtained by one call of separate routine
(see routines `{\tt spline}' to calculate second derivatives of
interpolating function and `{\tt splint}' to return a cubic spline
interpolated value in Press et al.$^{14}$ (1992)).

Our spline interpolation consists of the three steps:
\begin{itemize}
\item
      We set equidistant knots by the $\phi$--axis
      to reproduce equidistant grid;
\item
      We change grid by $x=\cos \theta $--axis to the
      required GLESP grid,
 \item
      after that, we recalculate $\phi$-knots to
      the rings corresponding to the GLESP $x$-points.
\end{itemize}

Fig. \ref{fig_hpgl_rep} demonstrates the deviation of accuracy
of the power spectrum
in a case of re-pixelization from a HEALPix map to a GLESP map with the
same resolution.
As one can see, for the range $\ell\le\ell_{max}/2$,
re-pixelization reproduces correctly all properties of the power spectra.
For $\ell\ge\ell_{max}/2$ some additional investigations needs to be done
to take into account the pixel-window function. This work is in progress.

\section{Resume}

We suggest a new scheme GLESP for sky pixelization based on the
Gauss--Legendre quadrature zeros. It has strict expansion by the
orthogonal functions which gives accuracy for $a_{\ell m}$--coefficients
calculations below $10^{-7}$ without any iterations.
We realized two approaches for Legendre polynomials calculation
using $L$-- and $M$--methods of calculation schemes.

Among the main advantages of this scheme are
\begin{itemize}
\item
      a high accuracy in calculation $a_{\ell m}$,
\item
      a high speed because of no iterations,
\item
      an optimal selection of resolution for a given beam size, which means
       an optimal number of pixels and a pixel size.
\end{itemize}

A corresponding code has been designed in FORTRAN\,77 and C languages
for procedures of the CMB sky map analysis.

The $a_{\ell m}$ calculation is the main goal.
$a_{\ell m}$-s  are used in component separation methods
and tests
for non-Gaussianity (Chiang et al.$^{10}$ 2003; Naselsky et al.$^{11,12}$ 2003b, 2004).
It is oriented on the fast and accurate calculation of the
$a_{\ell m}$ for the given resolution specified by the beam size.
Using accurately calculated $a_{\ell m}$-s,
one can reproduce any pixelization
scheme by the given pixel centers: GLESP, HEALPix, Igloo or Icosahedron.

\bigskip

{\bf Acknowledgments}.
This paper was supported in part by Danmark
Grundforskningsfond through its support for the establishment of the
Theorical Astrophysics Center.
Authors are thankful to Vladislav Stolyarov for testing parallel
capabilities and OPEN\,GL visualization tool for current pre-release version
of GLESP.
 OVV thanks the RFBR for partial support of the work through its grant
 02--07--90038.
Some of the results in this paper have been derived using
the HEALPix package (G\'orski, Hivon, and Wandelt$^{6}$ 1999).

\end{document}